\documentclass[prc, reprint, amsmath, amssymb, superscriptaddress, showpacs]{revtex4-1}

\usepackage{graphicx}%
\usepackage{multirow}
\usepackage{float}
\usepackage{epstopdf}
\usepackage{natbib}

\begin{document}

\title{Direct determination of the $^{138}$La $\beta$-decay $Q$ value using Penning trap mass spectrometry}%

\author{R. Sandler}%
\email{sandler@nscl.msu.edu}
\affiliation{Department of Physics, Central Michigan University, Mount Pleasant, Michigan, 48859, USA}
\affiliation{National Superconducting Cyclotron Laboratory, East Lansing, Michigan, 48824, USA}
\author{G. Bollen}
\affiliation{National Superconducting Cyclotron Laboratory, East Lansing, Michigan, 48824, USA}
\affiliation{Facility for Rare Isotope Beams, East Lansing, Michigan, 48824, USA}
\affiliation{Department of Physics and Astronomy, Michigan State University, East Lansing, Michigan 48824, USA}
\author{J. Dissanayake}
\affiliation{Department of Physics, Central Michigan University, Mount Pleasant, Michigan, 48859, USA}
\author{M. Eibach}
\affiliation{National Superconducting Cyclotron Laboratory, East Lansing, Michigan, 48824, USA}
\affiliation{Institut f\"ur Physik, Universit\"at Greifswald, 17487 Greifswald, Germany}
\author{K. Gulyuz}
\affiliation{Department of Physics, Central Michigan University, Mount Pleasant, Michigan, 48859, USA}
\author{A. Hamaker}
\affiliation{National Superconducting Cyclotron Laboratory, East Lansing, Michigan, 48824, USA}
\affiliation{Department of Physics and Astronomy, Michigan State University, East Lansing, Michigan 48824, USA}
\author{C. Izzo}
\affiliation{TRIUMF, Vancouver, British Colombia, Canada}
\author{X. Mougeot}
\affiliation{CEA, LIST, Laboratoire National Henri Becquerel (LNE-LNHB), B\^{a}t. 602 PC111, CEA-Saclay 91191 Gif-sur-Yvette Cedex, France.}
\author{D. Puentes}
\affiliation{National Superconducting Cyclotron Laboratory, East Lansing, Michigan, 48824, USA}
\affiliation{Department of Physics and Astronomy, Michigan State University, East Lansing, Michigan 48824, USA}
\author{F. G. A. Quarati}
\affiliation{AS, RST, LM, Delft University of Technology, Mekelweg 15, 2629JB Delft, The Netherlands}
\affiliation{Gonitec BV, Johannes Bildersstraat 60, 259EJ Den Haag, The Netherlands}
\author{M. Redshaw}
\affiliation{National Superconducting Cyclotron Laboratory, East Lansing, Michigan, 48824, USA}
\affiliation{Department of Physics, Central Michigan University, Mount Pleasant, Michigan, 48859, USA}
\author{R. Ringle}
\affiliation{National Superconducting Cyclotron Laboratory, East Lansing, Michigan, 48824, USA}
\author{I. Yandow}
\affiliation{National Superconducting Cyclotron Laboratory, East Lansing, Michigan, 48824, USA}
\affiliation{Department of Physics and Astronomy, Michigan State University, East Lansing, Michigan 48824, USA}
\date{\today}%

\begin{abstract}
\noindent
\begin{description}
\item[Background]
The understanding and description of forbidden decays provides interesting challenges for nuclear theory. These calculations could help to test underlying nuclear models and interpret experimental data.
\item[Purpose]
Compare a direct measurement of the $^{138}$La $\beta$-decay $Q$ value with the $\beta$-decay spectrum end-point energy measured by Quarati \textit{et al.} using LaBr$_{3}$ detectors [Appl. Radiat. Isot. \textbf{108}, 30 (2016)]. Use new precise measurements of the $^{138}$La $\beta$-decay and electron capture (EC) $Q$ values to improve theoretical calculations of the $\beta$-decay spectrum and EC probabilities.
\item[Method]
High-precision Penning trap mass spectrometry was used to measure cyclotron frequency ratios of $^{138}$La, $^{138}$Ce and $^{138}$Ba ions from which $\beta$-decay and EC $Q$ values for $^{138}$La were obtained. 
\item[Results]
The $^{138}$La $\beta$-decay and EC $Q$ values were measured to be $Q_{\beta}$ = 1052.42(41) keV and $Q_{EC}$ = 1748.41(34) keV, improving the precision compared to the values obtained in the most recent atomic mass evaluation [Wang, \textit{et al.}, Chin. Phys. C \textbf{41}, 030003 (2017)] by an order of magnitude. These results are used for improved calculations of the $^{138}$La $\beta$-decay shape factor and EC probabilities. New determinations for the $^{138}$Ce 2EC $Q$ value and the atomic masses of $^{138}$La, $^{138}$Ce, and $^{138}$Ba are also reported.
\item[Conclusion]
The $^{138}$La $\beta$-decay $Q$ value measured by Quarati \textit{et al.} is in excellent agreement with our new result, which is an order of magnitude more precise. Uncertainties in the shape factor calculations for $^{138}$La $\beta$-decay using our new $Q$ value are reduced by an order of magnitude. Uncertainties in the EC probability ratios are also reduced and show improved agreement with experimental data.
\end{description}
\end{abstract}

\maketitle

\section{Introduction}
Historically, nuclear $\beta$-decay studies have played a crucial role in our understanding of nuclear and particle physics and in the development of the Standard Model. Presently, high-precision and low-background nuclear $\beta$-decay experiments are being used to test the assumptions of the Standard Model and to search for new physics e.g. \cite{Har2015,Sev2006}. In addition to the exotic  neutrinoless double $\beta$-decay process \cite{Avi2008}, interest in other rare weak decay processes such as ultra-low $Q$ value $\beta$-decays \cite{Mus2010} and forbidden $\beta$-decays e.g. \cite{Mus2006,Haa2014,Gam2016,Kan2017}, has grown in recent years. The need for more precise $\beta$-spectrum shape measurements and calculations for forbidden $\beta$-decays is becoming apparent in a number of applications \cite{Kos2018}. For example, such input is necessary in the use of the proposed spectral shape method (SSM) to determine the effective value of the weak axial vector coupling constant, $g_{A}$ \cite{Haa2016}, and for understanding antineutrino spectra in context of the reactor antineutrino anomaly \cite{Hay2014,Son2015}. 

\begin{figure} [t]
\includegraphics[width=\columnwidth]{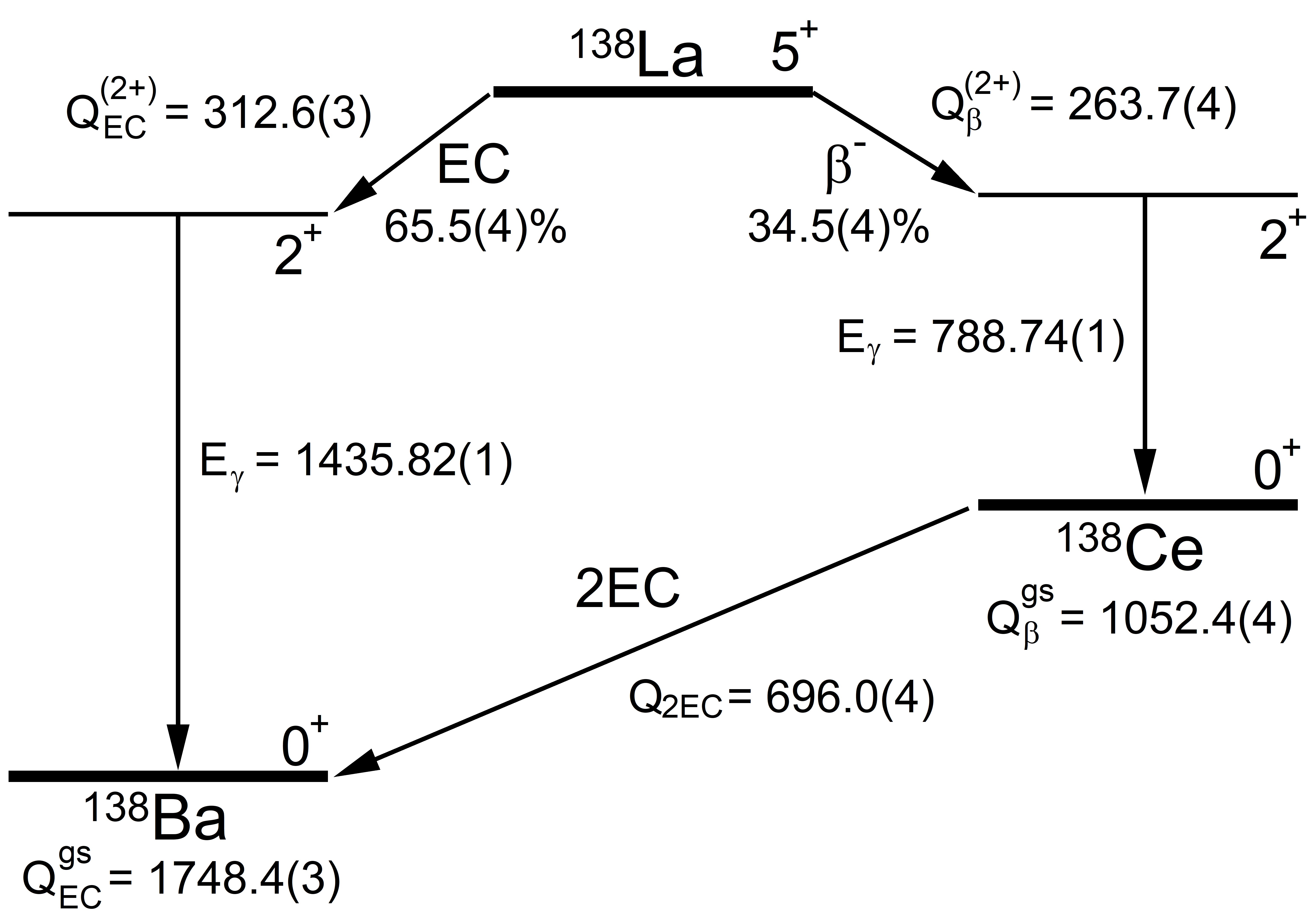}
\caption{Decay scheme for the Ba-La-Ce $A$ = 138 triplet. $Q_{\beta}^{gs}$ and $Q_{EC}^{gs}$ are the ground-state to ground-state $\beta$-decay and EC $Q$ values measured in this work, corresponding to the energy equivalent of the mass difference between parent and daughter atoms. $Q_{\beta}^{(2+)}$ and $Q_{EC}^{(2+)}$ are the $\beta$-decay and EC $Q$ values to the 2$^{+}$ daughter state in $^{138}$Ce and $^{138}$Ba, calculated as $Q^{(2+)}$ = $Q^{gs}$ -- E$_{\gamma}$. All $Q$ values and $\gamma$-energies are given in keV. \label{fig:Decay}}
\end{figure}

In this paper, we focus on the second forbidden unique decay of $^{138}$La. Naturally occurring $^{138}$La has a half-life of 1.03(1)$\times$10$^{11}$ years, and can undergo both $\beta^{-}$-decay to the 2$^{+}$ state in $^{138}$Ba and electron capture (EC) to the 2$^{+}$ state in $^{138}$Ce. In addition, $^{138}$Ce is energetically unstable against double EC to the $^{138}$Ba ground state. However, this decay has not been observed \cite{Bel2010}. A schematic of the decay scheme for this isobaric triplet system is shown in Fig.~\ref{fig:Decay}. 

Evidence for the radioactive decay of $^{138}$La was first obtained in 1950 \cite{Pri1950}, just a few years after its discovery \cite{Ing1947}. Since then, a series of measurements were performed that provided an understanding of the $^{138}$La decay scheme and more precise determinations of the partial and total half-lives \cite{Pri1951,Mul1952,Tur1956,Wat1962,DeR1966,Ell1972,Ces1977,Tay1979,Sat1981,Nor1983,Nir1997}. The long half-life has enabled the use of $^{138}$La for geochemical dating \cite{Bel2015} and as a nuclear cosmochronometer \cite{Hay2008}.

Recently, the development of LaBr$_{3}$ and LaCl$_{3}$ scintillation detectors has enabled new measurements of the $^{138}$La $\beta$-decay and EC x-ray spectra \cite{McI2011,Gia2015,qua2012,Qua2016}. From these measurements, more precise determinations of the relative EC probabilities and the $\beta$-decay spectrum shape can be made and compared with theoretical calculations. An experimental quantity that enters into these calculations is the $Q$ value for the decay, corresponding to the energy equivalent of the mass difference between the parent and daughter atoms, taking into account the energy of the daughter nuclear state. Before the $^{138}$La $\beta$-decay spectrum measurement by Quarati \it et al\rm. \cite{Qua2016}, the uncertainties in the relevant $Q$ values were limited by the uncertainties in the masses of $^{138}$La and $^{138}$Ce, as given in the 2012 atomic mass evaluation (AME2012) \cite{Aud2012}. The determination of the $^{138}$La $\beta$-decay spectrum end-point energy in Ref. \cite{Qua2016} reduced the uncertainty in the $^{138}$La $\beta$-decay and EC $Q$ values to 4.0 and 3.2 keV \cite{Aud2012}, respectively. In this paper, we present for the first time direct determinations of the $^{138}$La $\beta$-decay and EC $Q$ values using Penning trap mass spectrometry. We use these new $Q$ values to calculate EC ratios and $\beta$-spectrum shape factor coefficients. We also provide updated atomic masses for $^{138}$Ba, $^{138}$La, and $^{138}$Ce and for the $^{138}$Ce 2EC $Q$ value.

\section{Experimental Description}

\begin{figure} [h]
\includegraphics[scale=.23, trim = {1 1 1 1}, clip]{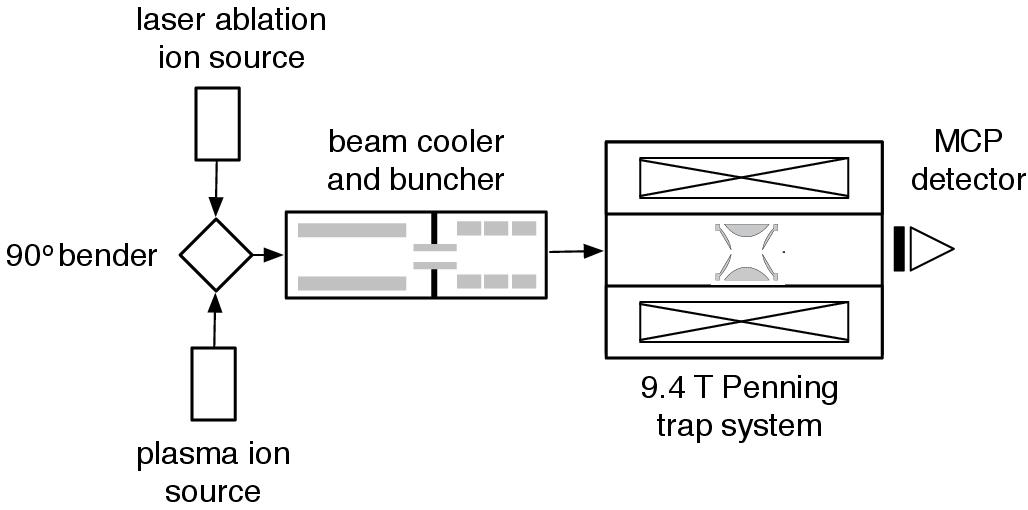}
\caption{A schematic overview of the sections of the LEBIT facility used for this experiment. \label{fig:Beamline}} 
\end{figure}

The $^{138}$La $\beta$-decay and EC $Q$ value measurements and absolute mass measurements were performed at the Low Energy Beam and Ion Trap (LEBIT) Penning trap mass spectrometry facility at the National Superconducting Cyclotron Laboratory (NSCL), a schematic of which is shown in Fig.~\ref{fig:Beamline}. LEBIT was designed for online measurements of rare isotopes from the Coupled Cyclotron Facility, but also houses two offline sources\textemdash a laser ablation source (LAS)~\cite{Izz2016} and a plasma ion source\textemdash that can be used for the production of stable and long-lived isotopes. These offline sources provide reference ions during rare isotope measurements, but also provide access to a wide range of isotopes that have been used for studies related to neutrinoless double $\beta$-decay~\cite{Red2012,Lin2013,Bus2013,Bus2013b,Gul2015,Eib2016}, highly forbidden $\beta$-decays~\cite{Gam2016,Kan2017}, and ultra-low $Q$ value $\beta$-decays~\cite{San2019}. 

The LAS, described in detail in~\cite{Izz2016}, uses a pulsed Nd:YAG laser to ablate material from a solid target. For this experiment, the LAS was fitted with 25 mm $\times$ 12.5 mm $\times$ 1 mm thick Ba, La, and Ce sheets of natural isotopic abundance~\cite{espi}. Two targets were placed on either side of the holder at one time and a stepper motor was used to alternate between the two sides. The high temperatures produced by the laser pulse results in the evaporation of surface material and the emission of positive ions and electrons to produce a high-temperature plasma. In addition to surface ionization, electron impact ionization of the ablated material, as well as other mechanisms contribute to the total ion production, see, e.g.~\cite{Sin1990} for a complete description. After production, ions are accelerated to an energy of 5 keV and focused into a 90 degree quadrupole bender that steers them into the main beamline. 

The plasma ion source is a DCIS-100 Colutron hot cathode discharge source~\cite{beam}. It consists of a tungsten filament within an alumina chamber. The chamber is filled with helium gas mixed with a small amount of xenon gas. As current is run through the filament it produces a discharge, creating a plasma within the gas-filled chamber. The ions are extracted through a radiofrequency quadrupole (RFQ) mass filter to suppress the helium ions, after which the xenon ions are focused into the other side of the 90 degree quadrupole bender and steered into the main beamline.

After entering the main beamline, ions are injected into an RFQ cooler and buncher \cite{Sch2016}. Helium buffer gas  is used to thermalize the ions, which are then released in packets of 100 ns duration to be accelerated to 2 keV and transported into the 9.4 T magnet containing the LEBIT Penning trap. At the entrance of the magnetic field is a fast electrostatic kicker, which only allows ions of the chosen $A/q$ to pass based on their time-of-flight. A series of electrodes decelerates the remaining ions to be captured in the Penning trap.  

The Penning trap itself consists of a hyperbolic ring electrode, two hyperbolic endcap electrodes, and two correction ring and correction tube electrodes that sit within a uniform magnetic field produced by a 9.4 T superconducting solenoidal magnet. The ring electrode of the Penning trap is segmented so that dipole and RFQ fields can be applied to address the radial modes of the ions' motion. Ions are confined radially in the trap via their cyclotron motion in the magnetic field that, without the presence of the electric field, has the frequency 
\begin{equation}
f_c = \frac{1}{2\pi}\frac{qB}{m}, \label{eq:Mass}
\end{equation}
where $B$ is the magnetic field strength and $m/q$ is the mass-to-charge ratio of the ion.

The trap electrodes produce a quadratic electrostatic potential that confines ions axially. The electric field also has the effect of reducing the frequency of the cyclotron motion of an ion and introducing an additional radial motion, the magnetron mode. As such, an ion in the Penning trap has three normal modes of motion: the axial, reduced cyclotron, and magnetron modes, with eigenfrequencies $f_z$, $f_+$ and $f_-$, respectively~\cite{Bro1986}. For an ideal Penning trap, the frequencies of the radial modes are related to the true cyclotron frequency of Eqn.~(\ref{eq:Mass})~\cite{Gab2008, Gab2009} via
\begin{equation}
f_++f_-=f_c. \label{eq:freq}
\end{equation}

\begin{figure}[h]
\includegraphics[width=\columnwidth]{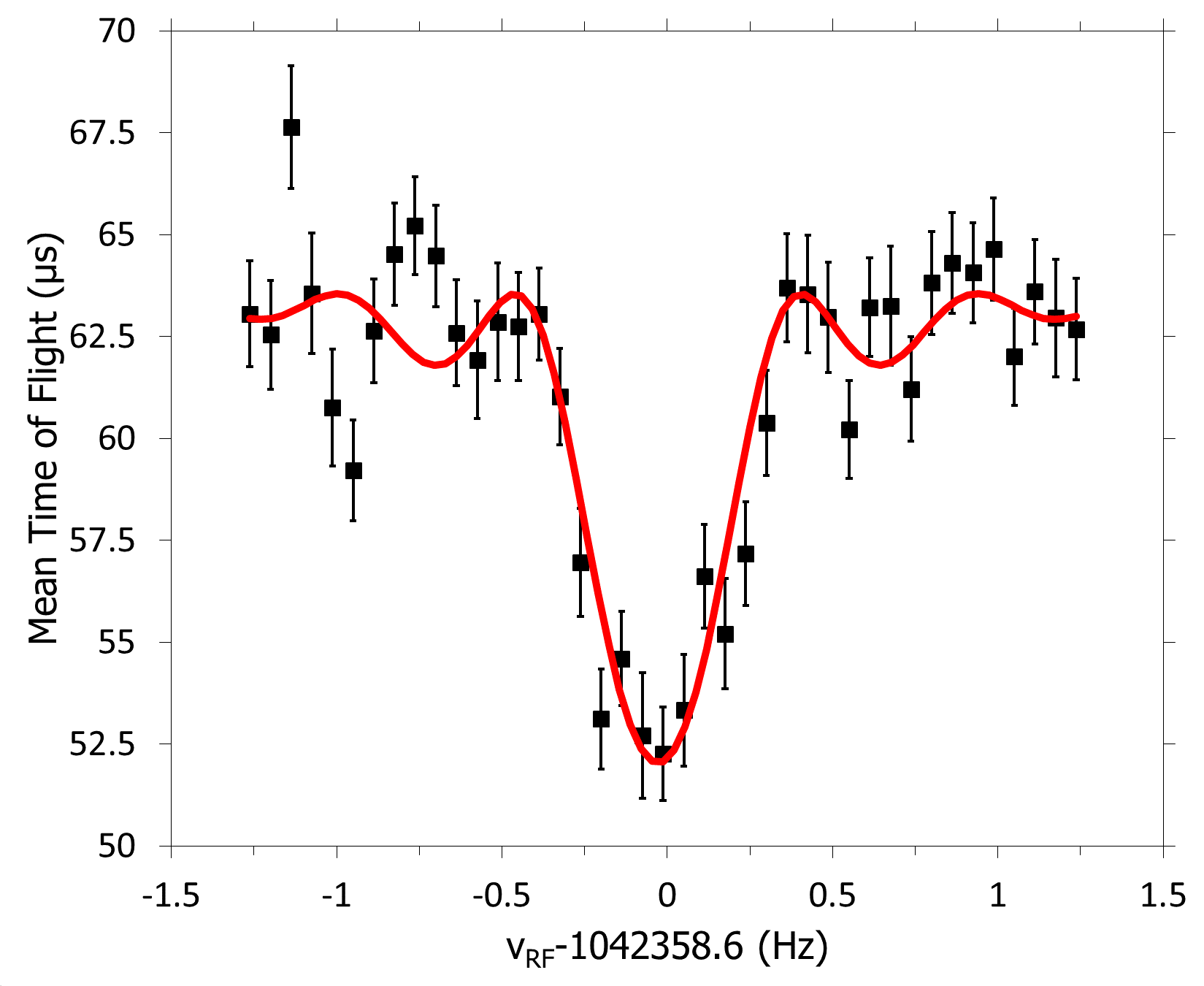}
\caption{A 2.0 s time-of-flight ion cyclotron resonance for $^{138}$La. The solid line is the theoretical fit to the data~\cite{Kon1995}. \label{fig:Resonance}} 
\end{figure}

Before entering the trap, ions are deflected off-axis by a Lorentz steerer and captured in a magnetron orbit of well-defined radius, typically $\sim$0.5 mm~\cite{rin2007}. A dipole RF pulse of 20 ms duration at the reduced cyclotron frequency of any previously identified contaminant ions is then applied to drive the contaminant ions into the trap walls~\cite{bol1996}. Next, the cyclotron frequency of the ion of interest is measured using the time-of-flight ion cyclotron resonance (TOF-ICR) technique~\cite{gra1980}. In this technique, an RFQ pulse of appropriate amplitude and duration is applied at the frequency $f_{RF} \approx f_{+} + f_{-}$. This pulse couples the reduced cyclotron and magnetron modes, which converts magnetron motion into cyclotron motion and increases the radial energy of the ions. The ions are then released from the trap and their time-of-flight to a microchannel plate (MCP) detector is recorded, which depends on the ions' initial radial energy. The measurement cycle is repeated over a range of values of $f_{RF}$ close to $f_{+} + f_{-}$ and a time-of-flight resonance curve such as the example shown in Fig.~\ref{fig:Resonance} is obtained. The minimum in time-of-flight corresponds to maximum radial energy, which results from a full conversion of magnetron to cyclotron motion by an RF pulse with $f_{RF} = f_{+} + f_{-} = f_{c}$. Hence, $f_c$ is obtained from a fit of the theoretical lineshape~\cite{Kon1995} to the data, as shown in Fig.~\ref{fig:Resonance}.

Our data taking procedure involved alternating between cyclotron frequency measurements on two ion species to account for temporal magnetic field variations. We measured $f_{c1}$ of ion 1 at time $t_1$, $f_{c2}$ of ion 2 at time $t_2$, and $f_{c1}$ of ion 1 again at time $t_3$. We then linearly interpolated the two $f_{c1}$ measurements to find the cyclotron frequency of ion 1 at time $t_2$. From this, we found the cyclotron frequency ratio, using the equation
\begin{equation}
R = \frac{f_{c1}(t_2)}{f_{c2}(t_2)} = \frac{m_2}{m_1}.\label{eq:R}
\end{equation}

\begin{figure} [h]
\includegraphics[width=\columnwidth]{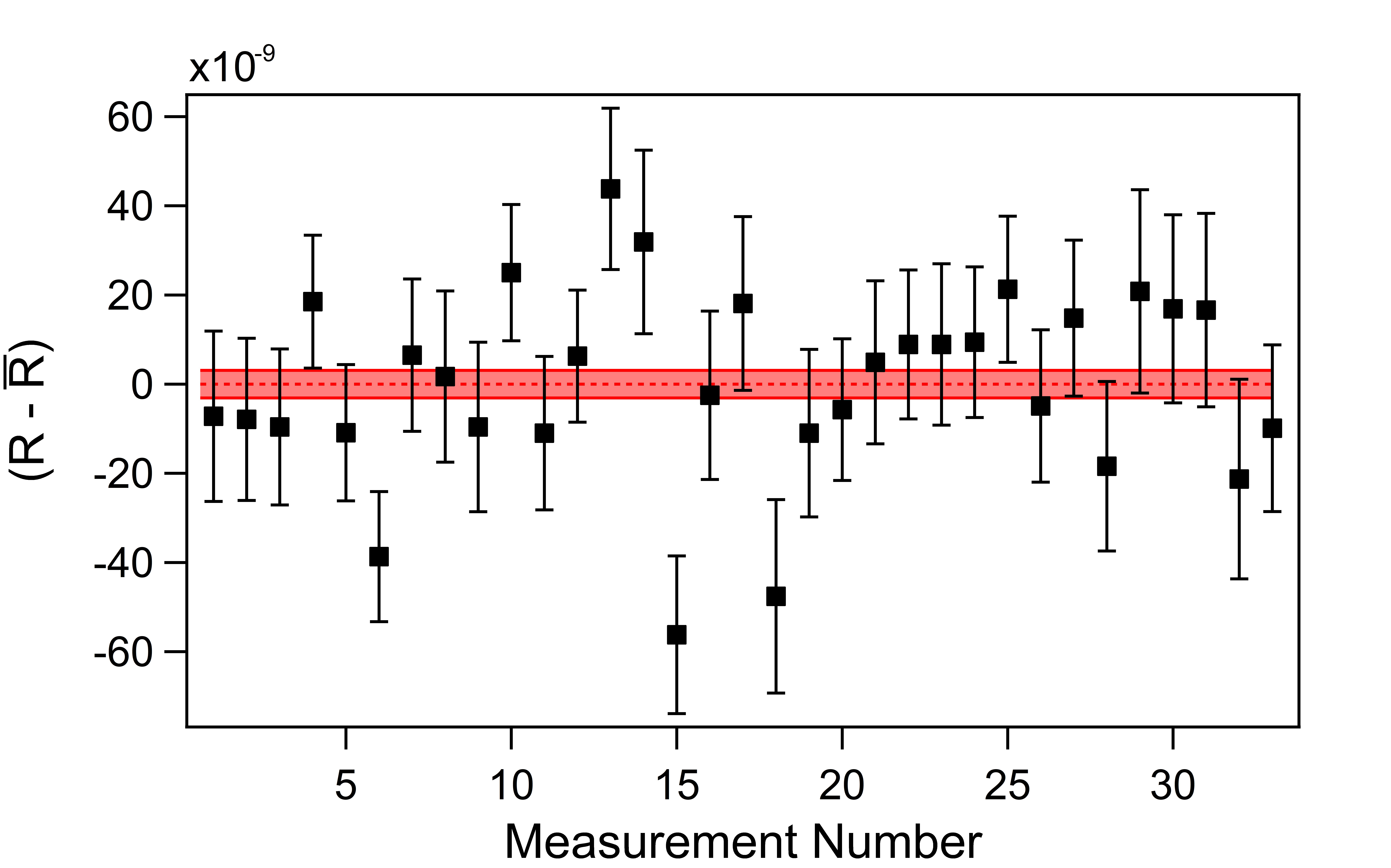}
\caption{Cyclotron frequency ratio measurements for $^{138}$La$^{+}$/$^{138}$Ce$^{+}$, with the 1$\sigma$ uncertainty in $\bar{R}$ shown by the shaded region.\label{fig:ratios}}
\end{figure}

We repeated this series of measurements twenty to fifty times and found the average cyclotron frequency ratio $\bar{R}$, as seen in Fig.~\ref{fig:ratios}. The Birge ratio~\cite{bir1932} for each series was calculated and, when the Birge Ratio was greater than 1, the uncertainty of $\bar{R}$ was inflated by the Birge ratio to account for possible underestimation of systematic uncertainty.

\section{Results and Discussion}
The cyclotron frequency ratios that we measured in this work, corresponding to inverse mass ratios of singly charged $^{138}$La,  $^{138}$Ce, $^{138}$Ba, and $^{136}$Xe ions, are given in Table~\ref{table:ratio}. 

\begin{table}[b]
\caption{\label{table:ratio} Measured cyclotron frequency ratios for combinations of $^{138}$La$^{+}$, $^{138}$Ba$^{+}$, and $^{138}$Ce$^{+}$ ions among themselves and against $^{136}$Xe$^{+}$. N is the number of individual ratio measurements contributing to the average, $\bar{R}$. The uncertainties for $\bar{R}$, shown in parentheses, have been inflated by the Birge Ratio, BR, when BR $>$ 1.}
\begin{ruledtabular}
\begin{tabular}{ccccc}
\multicolumn{1}{c}{Num.} & \multicolumn{1}{c}{Ion Pair} & \multicolumn{1}{c}{N} & \multicolumn{1}{c}{BR} & \multicolumn{1}{c}{$\bar{R}$} \\
\hline
(i) & $^{138}\text{La}^+$/$^{138}\text{Ce}^+$ & 33 & 1.2 & $0.999\ 991\ 810\ 7(37)$ \\
(ii) & $^{138}\text{La}^+$/$^{138}\text{Ba}^+$ & 48 & 1.1 & $0.999\ 986\ 387\ 2(29)$  \\
(iii) & $^{138}\text{Ce}^+$/$^{138}\text{Ba}^+$ & 32 & 1.0 & $0.999\ 994\ 589\ 6(56)$ \\
(iv) & $^{138}\text{Ce}^+$/$^{136}\text{Xe}^+$ & 79 & 1.3 & $0.985\ 506\ 162\ 7(118)$ \\
(v) & $^{138}\text{Ba}^+$/$^{136}\text{Xe}^+$ & 22 & 1.4 & $0.985\ 511\ 499\ 9(34)$  \\
\end{tabular}
\end{ruledtabular}
\end{table}

\subsection{$^{138}$La and $^{138}$Ce $Q$ value determinations}\label{sub:Qvalues}
The $\beta$-decay and EC $Q$ values are defined as the energy equivalent of the mass difference between parent and daughter atoms, $M_p$ and $M_d$, respectively. From this definition and Eqn.~(\ref{eq:R}), the $Q$ value for each decay can be obtained from the cyclotron frequency ratio measurement via
\begin{equation}
Q = (M_p - M_d)c^2 = (M_d-m_e)(1-\bar{R})c^2 \label{eq:Q},
\end{equation}
where $m_e$ is the mass of the electron and $c$ is the speed of light. Here we have ignored the ionization energies, which are nearly two orders of magnitude smaller than our statistical uncertainties and therefore do not affect our final results. The $Q$ values calculated using the cyclotron frequency ratios listed in Table~\ref{table:ratio} are given in Table~\ref{table:Q}. For each $Q$ value determination, we measured the relevant ratio in Eqn.~(\ref{eq:Q}) directly, e.g. ratio (i), $^{138}$La$^+$/$^{138}$Ce$^+$, is used to obtain $Q_{\beta}(^{138}$La$)$. However, we can also obtain the same ratio independently from the data in Table~\ref{table:ratio} from a ratio of ratios, e.g. (ii)/(iii) also gives $^{138}$La$^+$/$^{138}$Ce$^+$ where the intermediary nuclide is $^{138}$Ba. For each $Q$ value we list all such results and take the weighted average. 

\begin{table}[h]
\caption{\label{table:Q} $Q$ values for $^{138}$La $\beta$-decay or EC and $^{138}$Ce 2EC calculated from cyclotron frequency ratios listed in Table I. The relevant ratios were measured directly and via an intermediary isotope (see text for details). The weighted average is listed along with the AME2016 value \cite{Hua2017} and the difference $\Delta Q = Q_{\mathrm{LEBIT}} - Q_{\mathrm{AME}}$.}
\begin{ruledtabular}
\begin{tabular}{ccccc}
\multirow{2}{*}{Decay} & \multirow{2}{*}{Interm.} & \multicolumn{2}{c}{Q value (keV)} & \multicolumn{1}{c}{$\Delta Q$} \\
 &  & \multicolumn{1}{c}{LEBIT} & AME2016 & (keV)\\
 \hline
\multirow{3}{*}{$^{138}$La$\rightarrow^{138}$Ce} & Direct & 1051.98(48) & \multirow{3}{*}{} & \multirow{3}{*}{} \\
 & $^{138}$Ba & 1053.67(81) & & \\
($\beta$-) & Avg. & 1052.42(41) & 1051.7(4.0) & 0.7(4.0)\\
\hline
\multirow{3}{*}{$^{138}$La$\rightarrow^{138}$Ba} & Direct & 1748.67(37) & \multirow{3}{*}{} & \multirow{3}{*}{} \\
 & $^{138}$Ce & 1746.98(86) & & \\
(EC) & Avg. & 1748.41(34) & 1742.5(3.2) & 5.9(3.2)\\
\hline
\multirow{4}{*}{$^{138}$Ce$\rightarrow^{138}$Ba} & Direct & \ 695.01(72) & \multirow{4}{*}{}  & \multirow{4}{*}{} \\
 & $^{138}$La & \ 695.68(1.58) & & \\
 & $^{136}$Xe  & \ 696.69(60) & & \\
(2EC) & Avg. & \ 695.97(44) & 690.7(4.9) & 5.3(4.9) \\                                         
\end{tabular}
\end{ruledtabular}
\end{table}

\subsubsection{$^{138}$La $\beta$-decay $Q$ value} 
\begin{figure}[b]
\includegraphics[width=\columnwidth]{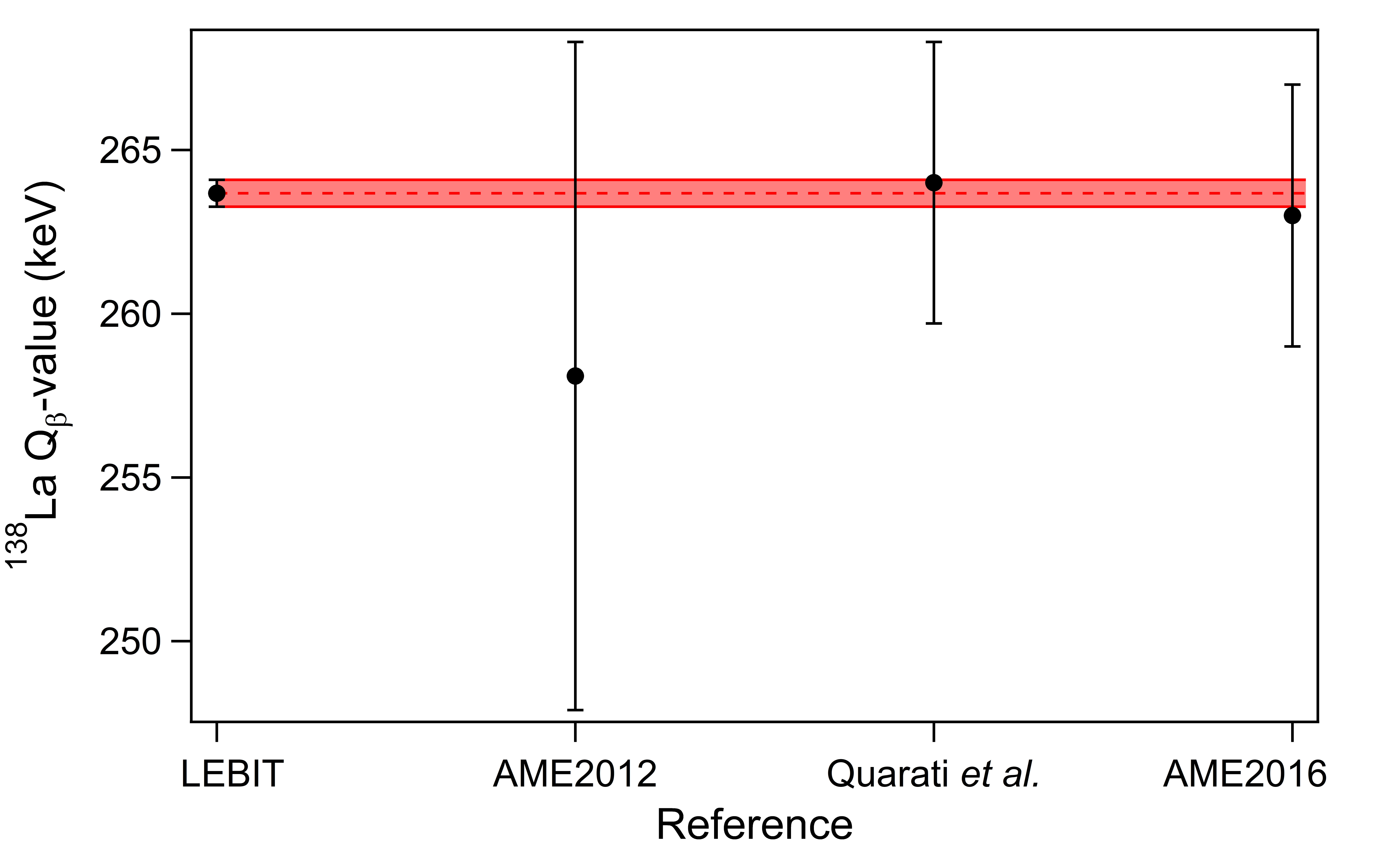}
\caption{LEBIT $^{138}$La $\beta$-decay $Q$ value result compared to the AME2012 \cite{Aud2012}, AME2016 \cite{Hua2017} and Quarati, \textit{et al.} \cite{Qua2016} values. \label{fig:Results}} \end{figure}

One of the main motivations of this work was to perform a precise measurement of the $^{138}$La $\beta$-decay $Q$ value using Penning trap mass spectrometry to compare with the result of Quarati \textit{et al.}~\cite{qua2012} obtained from a measurement of the end-point energy of the $^{138}$La $\beta$-decay spectrum using LaBr$_3$ detectors. A comparison of these results can be seen in Fig.~\ref{fig:Results} along with results from the AME2012 and AME2016 \cite{Aud2012,Hua2017} (we note that the AME2016 analysis includes the Quarati \textit{et al.} result). For this comparison, we compute the $\beta$-decay spectrum end-point energy, corresponding to the $Q$ value defined in Eqn.~(\ref{eq:Q}) with the energy of the $^{138}$Ce(2+, 788.74 keV) daughter state subtracted. The Quarati \textit{et al.} result of 264.0(4.3) keV is in excellent agreement with our new value of 263.68(41) keV, which is an order of magnitude more precise.

\subsubsection{$^{138}$La EC $Q$ value determination}
Our direct measurement of the $^{138}$La EC $Q$ value shows a 5.9(3.2) keV shift with respect to the AME2016 value and a reduction in uncertainty of almost an order of magnitude. Our direct mass determinations of $^{138}$Ce and $^{138}$Ba, described in section~\ref{sub:masses}, indicate that this disagreement is due to a shift in the mass of $^{138}$Ce compared to the AME2016 value. Since the mass of $^{138}$La is directly linked to the mass of $^{138}$Ce in the AME2016 through the $^{138}$La $\beta$-decay $Q$ value measurement of Quarati, \textit{et al.}, \cite{Qua2016} the $^{138}$La mass is also shifted with respect to the AME2016 value. Our new measurement enables more precise calculations of the $^{138}$La relative EC probabilities, as described in section~\ref{sub:calculations}.

\begin{figure} [t]
\includegraphics[width=\columnwidth]{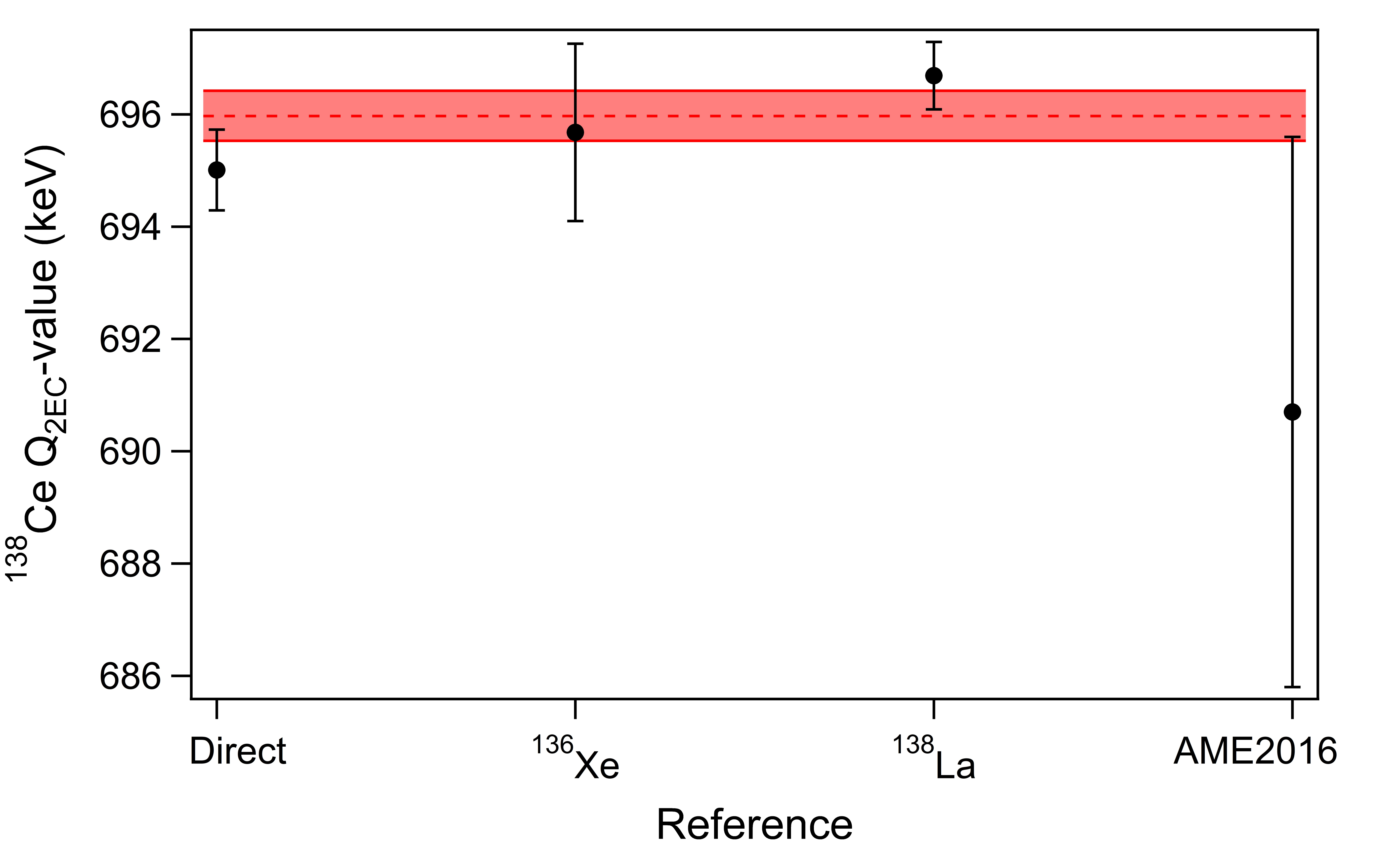}
\caption{LEBIT $^{138}$Ce 2EC $Q$ value measurements and their weighted average and uncertainty (shown by the dotted line and shaded region) compared to the AME2016 value. \label{fig:Q2EC}} 
\end{figure}

\subsubsection{$^{138}$Ce 2EC $Q$ value}
Finally, in Table~\ref{table:Q}, we list three independent results for the $^{138}$Ce $Q_{2EC}$-value along with their weighted average. The first result is a direct measurement of the $Q$ value obtained from ratio (iii) in Table~\ref{table:ratio}, using Eqn.~(\ref{eq:Q}). The second and third results are from the ratio of ratios of (ii)/(i) and (iv)/(v), respectively using $^{138}$La and $^{136}$Xe as an intermediary. These results and their weighted average are plotted in Fig.~\ref{fig:Q2EC} along with the value obtained from the AME2016. Our three independent measurements of the $^{138}$Ce $Q_{2EC}$-value are in good agreement with each other, but the average shows a 5.3(4.9) keV discrepancy with respect to the AME2016 value. Again, our direct mass determinations of $^{138}$Ce and $^{138}$Ba, described in section~\ref{sub:masses}, indicate that this disagreement is due to a shift in the mass of $^{138}$Ce compared to the AME2016 value.


\subsection{$^{138}$La $\beta$-spectrum shape factor and $EC$ ratio calculations}\label{sub:calculations}

It has been well known for a long time that the mass region around $^{138}$La cannot be depicted by a naive shell model~\cite{Helton1973} and that the collective structure of the nuclear states is critical to reproduce low energy data ~\cite{Suhonen1993}. In this context, precise measurements are of high importance to test and constrain nuclear models. In this section, we study the influence of a precise knowledge of $Q$ values on the theoretical predictions. We first look at the electron energy spectrum from the $\beta$-decay to $^{138}$Ce and then at the capture probabilities from the EC decay to $^{138}$Ba. 

\subsubsection{$^{138}$La $\beta$-spectrum shape factor}

The $\beta$-decay spectrum can be described, following the formalism of Behrens and B\"{u}ring~\cite{Behrens82}, as
\begin{equation}
\frac{dN}{dW} \propto pWq^{2}F_{0}L_{0}C(W),\label{eq:BetaSpec}
\end{equation}
where $W$ is the total electron energy, $p$ its momentum and $q$ the antineutrino energy. The Fermi function $F_{0}L_{0}$ is defined from the Coulomb amplitudes of the relativistic electron wave functions which are solutions of the Dirac equation for a static Coulomb potential from a uniformly charged sphere. The theoretical shape factor $C(W)$ couples the nuclear structure of the nuclei involved in the decay with the lepton dynamics. Describing the weak interaction as a current-current interaction, a multipole expansion can be performed for each current \textemdash the hadron current and the lepton current. Keeping only the main terms, the nuclear component can be factored out of the theoretical shape factor for allowed and forbidden unique transitions. In the present work, we have calculated the second forbidden unique transition from the ground-state of $^{138}$La to the first excited state of $^{138}$Ce, for which one has:
\begin{equation}
C(W) = q^{4} + \frac{10}{3} \lambda_{2} q^{2} p^{2} + \lambda_{3} p^{4},\label{eq:ShapeFactor}
\end{equation}
where the $\lambda_{k}$ parameters are ratios of Coulomb amplitudes of the electron wave functions. 

\begin{figure}
\includegraphics[scale=0.47]{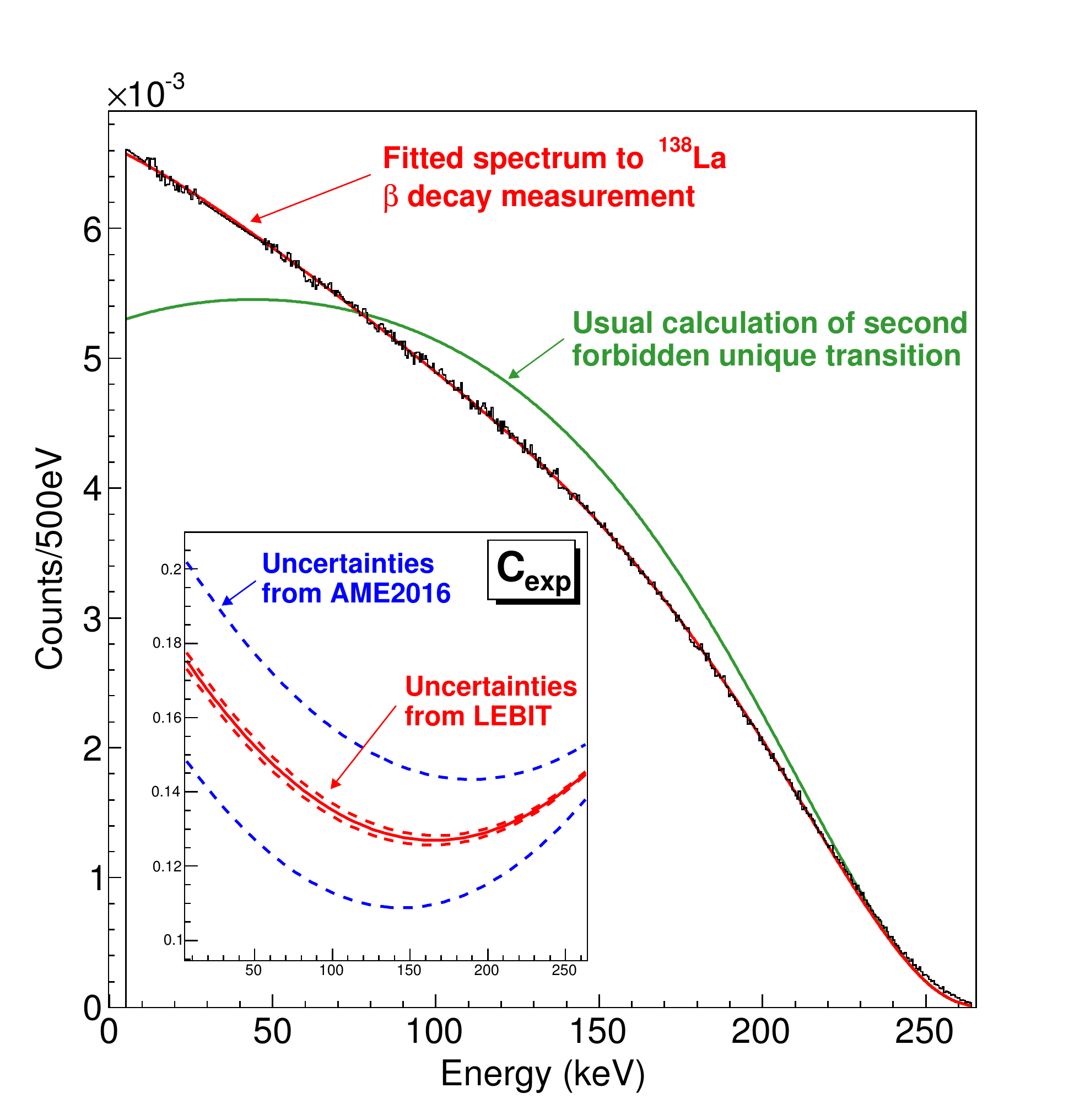}
\caption{Extraction of experimental shape factor $C_{\text{exp}}$ for $^{138}$La $\beta$-decay using AME2016 $Q$ value from Ref. \cite{Hua2017} and LEBIT $Q$ value from this work. The measured spectrum, shown in black, is from Ref. \cite{Qua2016}. The classical theoretical calculation is shown in green. $C_{\text{exp}}$ is applied to the theory to get the adjusted spectrum, shown in red. The inset shows the improvement on $C_{\text{exp}}$ uncertainties due to the high-precision LEBIT $Q$ value determination. \label{fig:BetaSpec}}
\end{figure}

This treatment of the shape factor usually gives good agreement with measurements~\cite{Mougeot2015}. However, $^{138}$La exhibits a specific nuclear structure which leads to an accidental cancellation of the nuclear matrix elements. The leading multipole orders are not sufficient anymore to describe the transition and higher orders have to be included. This mechanism hinders the transition and drastically increases the half-life. As shown in Fig.~\ref{fig:BetaSpec}, it also modifies the shape of the $\beta$ spectrum, our calculation (green) being far from the measured spectrum (black) from Ref.~\cite{Qua2016}. Therefore, we have performed fits to these data to determine an experimental shape factor $C_{\text{exp}}$ defined as the distortion to be applied on the theoretical shape factor to get the measured spectrum. A minimum of two parameters was necessary to fit the data, with the form $C_{\text{exp}}(W) = 1 + aW + bW^{2}$. For these fits we used an end-point energy, $E_{\text{max}}$, of either 263.3(4.0) keV obtained from the AME2016~\cite{Hua2017}, or 263.68(41) keV found in this work. Uncertainty limits on the parameters were determined by refitting the data with $E_{\text{max}} \rightarrow E_{\text{max}} \pm \sigma_{E_{\text{max}}}$. The resulting parameters and corresponding uncertainties are shown in Table~\ref{table:BetaSpec} and are illustrated in Fig.~\ref{fig:BetaSpec}. As can be seen, the results are very consistent and the new $Q$ value reduces uncertainties in the shape factor fit parameters by a factor of $\sim 11$, putting a stronger constraint on the precision of future predictions of the nuclear matrix elements. 

\begin{table}[h]
\caption{\label{table:BetaSpec} Adjusted parameters of the experimental shape factor $C_{\text{exp}}(W) = 1 + aW + bW^{2}$, to be applied on the theoretical shape factor Eqn.~(\ref{eq:ShapeFactor}) to match the measured spectrum from Ref.~\cite{Qua2016}. The fitting procedure has been applied using the AME2016 $Q$ value from Ref.~\cite{Hua2017} and the LEBIT $Q$ value from this work. Upper uncertainties are for $E_{\text{max}} + \sigma_{E_{\text{max}}}$ and lower uncertainties for $E_{\text{max}} - \sigma_{E_{\text{max}}}$.}
\begin{ruledtabular}
\begin{tabular}{ccccc}
\multirow{2}{*}{Parameter} & \multicolumn{2}{c}{AME2016} & \multicolumn{2}{c}{LEBIT}\\
 & value & uncertainty & value & uncertainty\\
\hline
\multirow{2}{*}{a} & \multirow{2}{*}{$-$1.32} & +0.07 & \multirow{2}{*}{$-$1.319} & +0.006 \\
  &  & $-$0.07 & & $-$0.006 \\
\hline
\multirow{2}{*}{b} & \multirow{2}{*}{0.499} & $-$0.043 & \multirow{2}{*}{0.4982} & $-$0.0038 \\
  &  & $+$0.043 & & +0.0038 \\
\hline
$\chi^{2}$ & \multicolumn{2}{c}{9.0$\times$10$^{-5}$} & \multicolumn{2}{c}{8.8$\times$10$^{-5}$}\\
\end{tabular}
\end{ruledtabular}
\end{table}

\subsubsection{$^{138}$La $EC$ ratio calculations}
We have performed the calculation of the second forbidden unique electron capture transition from the ground-state of $^{138}$La to the first excited state of $^{138}$Ba. The modeling used has already been described in Ref.~\cite{Mou2018} and takes into account overlap, exchange, shake-up and shake-off, and hole effects. However, radiative corrections based on Coulomb-free theory~\cite{Bambynek1977} have also been considered in the present work. In addition, the relativistic atomic wave functions were determined using the precise atomic orbital energies from Refs.~\cite{Koto1997,KotoErr1997} which include the effect of electron correlations. The resulting EC probability ratios for $K$, $L$, and $M$ shells are shown in Table~\ref{table:ECRatios}. The calculations were performed using $Q_{EC}$ = 1742(3) keV from the AME2016~\cite{Hua2017} and $Q_{EC}$ = 1748.41(34) keV obtained in this work and are compared with the precise measurements from Ref.~\cite{Qua2016}. A reduction in the uncertainties of the calculated values by factors of 2.4 to 3 is achieved with the new $Q$ value. It is noteworthy that a change of the $Q$ value by less than 0.4\% leads to a perfect agreement of the predicted $L/K$ value with the measured one. The differences between predictions and measurements for the $M/K$ and $M/L$ values can be explained by the low energies of the $M$ subshells, which make both their high-precision calculation and measurement very difficult. 

The calculations shown in Table~\ref{table:ECRatios} have been performed following the usual approximation of a constant nuclear component, identical for each subshell, which cancels when looking at capture probability ratios. This assumption is considered to be correct for both allowed and forbidden unique transitions~\cite{Bambynek1977}. However, in order to investigate the sensitivity of our theoretical predictions to the inclusion of the nuclear component, besides that reported in Table~\ref{table:ECRatios}, we have performed additional calculations of the capture probability ratios. We have followed the formalism of Behrens and B\"{u}ring~\cite{Behrens82} in which, as for $\beta$-decays, the coupling of the nuclear and lepton components is given for each subshell through a double multipole expansion by:
\begin{equation}
C_{\kappa_x} = \sum\limits_{K,\kappa_{\nu}}{[ M_K(\kappa_x,\kappa_{\nu}) + S_{\kappa_x} m_K(\kappa_x,\kappa_{\nu}) ]^{2}},\label{eq:CxFactor}
\end{equation}
where $\kappa_x$ and $\kappa_{\nu}$ are quantum numbers of the electron and neutrino respectively, and $S_{\kappa_x}$ is the sign of $\kappa_x$. The $M_K$ and $m_K$ quantities include nuclear and lepton matrix elements. They have been determined in impulse approximation considering the single decay of a $1g_{7/2}$ proton in $^{138}$La to a $3s_{1/2}$ neutron in $^{138}$Ba. A non-relativistic harmonic oscillator modeling has been considered for the large component of the relativistic nucleon wave functions, and the small component has been estimated following the method given in Ref. \cite{Behrens82}. With the $Q$ value from this work, we found a significant change in the $L/K$ ratio by taking into account the nuclear component\textemdash $L/K_{nuc} = 0.3827(26)$\textemdash while the other two capture probabilities remain consistent\textemdash $M/K_{nuc} = 0.0962(10)$, and $M/L_{nuc} = 0.2514(31)$. One can clearly see that a high-precision determination of the $Q$ value allows for testing of the accuracy of the nuclear model, eventually providing nuclear structure information. A more realistic treatment would necessitate taking into account nucleus deformation and configuration mixing.

\begin{table}
\caption{\label{table:ECRatios} Influence of the $Q$ value on the theoretical predictions of the capture probability ratios for $^{138}$La. Experimental values are from Ref.~\cite{Qua2016}. The AME2016 $Q$ value is 1742(3) keV from Ref.~\cite{Hua2017} and the LEBIT $Q$ value is 1748.41(34) keV from this work.}
\begin{ruledtabular}
\begin{tabular}{cccc}
EC Ratio & Experiment & AME2016 & LEBIT\\
\hline
L/K & 0.391(3) & 0.403(8) & 0.3913(26)\\
M/K & 0.102(3) & 0.0996(24) & 0.0964(10)\\
M/L & 0.261(9) & 0.247(8) & 0.2464(30)\\
\end{tabular}
\end{ruledtabular}
\end{table}


\subsection{$^{138}$La, $^{138}$Ce, $^{138}$Ba atomic mass determinations}\label{sub:masses}
\begin{figure} [t]
\includegraphics[width=\columnwidth]{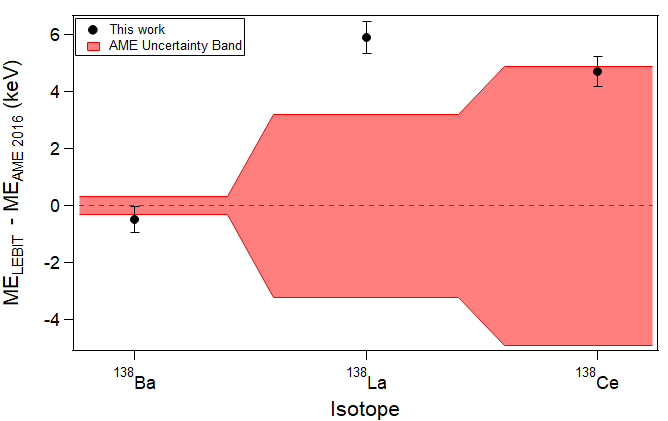}
\caption{Mass excesses measured in this work, as listed in Table~\ref{table:mass}, and compared to the AME2016 values, with AME2016 uncertainties indicated by the shaded region.\label{fig:masses}} 
\end{figure}

The absolute masses of $^{138}$La, $^{138}$Ce and $^{138}$Ba were obtained from our cyclotron frequency ratio measurements listed in Table~\ref{table:ratio} and the relation
\begin{equation}
M_{int}= (M_{ref}-m_e)\frac{1}{\bar{R}}+m_e,\label{eq:RMass}
\end{equation}
where $M_{int}$ and $M_{ref}$ are the atomic masses of the nuclide of interest and reference nuclide, respectively. Ratio (v) in Table~\ref{table:ratio}, $^{138}$Ba$^+$/$^{136}$Xe$^+$, provided a direct link to obtain the mass of $^{138}$Ba using $^{136}$Xe as a reference, which has been measured to a precision of 0.007 keV using the Florida State University Penning trap~\cite{Red2007}. We then used $^{138}$Ba as a secondary mass reference along with ratios (ii) and (iii) from Table~\ref{table:ratio} to obtain atomic masses for $^{138}$La and $^{138}$Ce respectively. Ratio (iv) in Table~\ref{table:ratio}, $^{138}$Ce$^+$/$^{136}$Xe$^+$, provided an independent check for the mass of $^{138}$Ce. The two results for $^{138}$Ce are in good agreement, although the second is a factor of two less precise. This was due to the fact that after operating the LAS with barium, it became contaminated and a background of $^{138}$Ba$^+$ was produced along with $^{138}$Ce$^+$. Finally, $^{138}$La was used as a secondary mass reference along with ratio (i) in Table~\ref{table:ratio} to calculate a third atomic mass. The three values of $^{138}$Ce are in good agreement and were used to calculate an average value for the atomic mass. The resulting masses excesses for $^{138}$Ba, $^{138}$La, and $^{138}$Ce are listed in Table~\ref{table:mass} and plotted in Fig.~\ref{fig:masses}.

Our result for the mass of $^{138}$Ba is in good agreement with the AME2016 value, which was determined from ($n,\gamma$) measurements along the barium isotope chain, a $^{134}$Cs $\rightarrow$ $^{134}$Ba $\beta$-decay measurement, a $^{133}$Cs($n,\gamma$)$^{134}$Cs measurement, and a Penning trap measurement of $^{136}$Ba$^{+}$/$^{136}$Xe$^{+}$ \cite{Kol2011}. These measurements anchor $^{138}$Ba to $^{133}$Cs \cite{Bra1999,Mou2010} and $^{136}$Xe \cite{Red2007}, which have been precisely measured with Penning traps and can be considered secondary mass standards. 

The determination of the masses of $^{138}$La and $^{138}$Ce in the AME is more convoluted. The mass of $^{138}$Ce is determined almost entirely from the Quarati \textit{et al.} $\beta$-decay end-point energy measurement and the mass of $^{138}$La. The mass of $^{138}$La on the other hand is partially obtained from a $^{138}$La($d,p$)$^{139}$La reaction measurement, and a $^{139}$Ba $\rightarrow$ $^{139}$La $\beta$-decay measurement that link it to the barium isotopes and ultimately $^{133}$Cs and $^{136}$Xe, as discussed above. It is also partially determined from a network of neutron capture, $\beta$-decay and $\alpha$-decay measurements that link the lanthanides up to $^{163}$Dy and $^{163}$Ho for which precise Penning trap measurements have been performed \cite{Eli2015}. Our results, listed in Table~\ref{table:mass} and displayed in Fig.~\ref{fig:masses}, indicate a discrepancy in the AME2016 mass values for both $^{138}$La and $^{138}$Ce of about 5 keV/c$^{2}$.

\begin{table}[b]
\caption{\label{table:mass} Mass excesses, ME, for $^{138}$Ba, $^{138}$La, and $^{138}$Ce obtained from the ratios listed in Table~\ref{table:ratio}. The results are compared to those listed in the AME2016 \cite{Hua2017}. The column $\Delta$M is calculated as ME$_{\rm{LEBIT}}$ -- ME$_{\rm{AME2016}}$}
\begin{ruledtabular}
\begin{tabular}{ccccc}
\multirow{2}{*}{Nuclide} & \multirow{2}{*}{Ref.} & \multicolumn{2}{c}{ME (keV/c$^2$)} & \multicolumn{1}{c}{$\Delta$M}\\
 & & LEBIT & \multicolumn{1}{c}{AME2016} & \multicolumn{1}{c}{(keV/c$^2$)}\\
 \hline
$^{138}$Ba & $^{136}$Xe & -88\ 262.13(0.44) & -88\ 261.64(0.32) & -0.49(0.54) \\
 \hline
$^{138}$La & $^{138}$Ba & -86\ 513.44(0.57) & -86\ 519.2(3.2) & \hspace{0.1em} 5.8(3.2)\\
 \hline
 & $^{138}$Ba & -87\ 567.12(0.84) & & \\
 & $^{136}$Xe & -87\ 566.45(1.54) & & \\
 & $^{138}$La & -87\ 565.43(0.74) & & \\ 
 $^{138}$Ce & Avg. & -87\ 566.21(0.52) & -87\ 570.9(4.9) & \hspace{0.1em} 4.7(4.9)\\
\end{tabular}
\end{ruledtabular}
\end{table}

As a check of possible systematics we performed a measurement of the mass ratios of $^{134}$Xe$^+$/$^{136}$Xe$^+$ and $^{136}$Ba$^+$/$^{136}$Xe$^+$, with the results $\bar{R}$ = 0.985 270 617 0(22) and 0.999 980 585 7(23) respectively. The ratios differ from those calculated using the AME2016 mass values for $^{134,136}$Xe and $^{136}$Ba, and $m_{e}$ = 5.485 799 090 70(16) $\times$ 10$^{-4}$ u \cite{Moh2016} by only $-$0.8(2.2) and 0.1(3.3) $\times$ 10$^{-9}$, respectively. This is well within acceptable deviation and is considered consistent with the AME.

\section{Conclusion}

Using Penning trap mass spectrometry, we have measured the $Q_{\beta}$-value of $^{138}$La to be 1052.42(41) keV and the $Q_{EC}$-value of $^{138}$La to be 1748.41(34) keV. Both measurements reduce the uncertainties compared to previous values by an order of magnitude. The determination of the $^{138}$La $\beta$-decay $Q$ value from a measurement of the end-point energy of the $\beta$-spectrum obtained with LaBr$_{3}$ detectors by Quarati, \textit{et al.}~\cite{Qua2016} is in excellent agreement with our new, more precise result. 

We have used our new $Q_{\beta}$ value in theoretical fits to the data of Ref. \cite{Qua2016} and extracted new values for the experimental shape factor parameters with uncertainties that are reduced by about an order of magnitude compared to those obtained using the $Q$ value from the AME2016. We have used our new $Q_{EC}$ value in theoretical calculations of the EC probabilities that we compare with the experimental EC ratio results of Ref. \cite{Qua2016}. Our new $Q$ value reduces the uncertainties in the calculated ratios by factors of up to 3 compared calculations using the $Q$ value from AME2016, and, for the case of the L/K ratio significantly improves the agreement between experiment and theory.

Finally, we also present the first direct mass measurements of $^{138}$La, $^{138}$Ce, and $^{138}$Ba. Our result for $^{138}$Ba is in good agreement with the AME2016 value with a similar level of precision. Our results for $^{138}$La and $^{138}$Ce show an $\approx$5 keV/c$^{2}$ shift with respect to AME2016 and reduce the uncertainties by factors of 6 and 9, respectively.

\section*{Acknowledgments}

This research was supported by Michigan State University and the Facility for Rare Isotope Beams and the National Science Foundation under Contracts No. PHY-1102511 and No. PHY1307233. This material is based upon work supported by the US Department of Energy, Office of Science, Office of Nuclear Physics under Award No. DE-SC0015927. The work leading to this publication has also been supported by a DAAD P.R.I.M.E. fellowship with funding from the German Federal Ministry of Education and Research and the People Programme (Marie Curie Actions) of the European Union’s Seventh Framework Programme (FP7/2007/2013) under REA Grant Agreement No. 605728. 


The theoretical work was performed as part of the EMPIR Projects 15SIB10 MetroBeta and 17FUN02 MetroMMC. These two projects have received funding from the EMPIR programme co-financed by the participating states and from the European Union's Horizon 2020 research and innovation programme.



\bibliography{La_paper} 


\end{document}